\documentclass{article}
\usepackage{spconf,amsmath,graphicx}

\usepackage{times}
\usepackage{tabularx}
\usepackage{amssymb}
\usepackage{amsthm}
\usepackage{floatrow}
\usepackage{caption}
\usepackage{subcaption}
\usepackage{capt-of, blindtext}
\usepackage{setspace}
\usepackage{array}
\usepackage{multirow}
\usepackage{fmtcount}
\usepackage{psfrag}
\usepackage{tikz}
\usepackage{ifthen}
\usepackage[pagebackref=true,breaklinks=true,letterpaper=true,colorlinks,bookmarks=false]{hyperref}
\usepackage{pbox}
\usepackage{hyperref}

\graphicspath{ {.} }

\newfloatcommand{capbtabbox}{table}[][\FBwidth]
\newfloatcommand{capbfigbox}{figure}[][\FBwidth]

\title{Detection of speech events and speaker characteristics \\ through photo-plethysmographic signal neural processing}
\name{Guillermo C\'ambara$^{1,2}$, Jordi Luque$^{1,3}$ and Mireia Farr\'us$^2$}
\address{$^1$Telefonica Research, Barcelona, Spain  \\
$^2$Universitat Pompeu Fabra, Barcelona, Spain \\
$^3$Universitat Polit\`ecnica de Catalunya, Barcelona, Spain \\}

\begin{document}
%
\maketitle
\begin{abstract}
The use of photoplethysmogram signal (PPG) for heart and sleep monitoring is commonly found nowadays in smartphones and wrist wearables. Besides common usages, it has been proposed and reported that person information can be extracted from PPG for other uses, like biometry tasks. In this 
work, we explore several end-to-end convolutional neural network 
architectures for detection of human's characteristics such as gender or person identity. In addition, we evaluate whether speech/non-speech events may be inferred from PPG signal, where speech might translate in fluctuations into the pulse signal. 
The obtained results are promising and clearly show the potential of fully end-to-end topologies for automatic extraction of meaningful biomarkers, even from a noisy signal sampled by a low-cost PPG sensor. The AUCs for best architectures put forward PPG wave as biological discriminant, reaching $79\%$ and $89.0\%$, respectively for gender and person verification tasks. Furthermore, speech detection experiments reporting  AUCs around $69\%$ encourage us for further exploration about the feasibility of PPG for speech processing tasks.

\end{abstract}
\begin{keywords}
Photoplethysmogram signal, PPG, speech detection, convolutional neural networks, biometric authentication
\end{keywords}
\section{Introduction}
\label{sec:intro}

Nowadays, tens of biometric sensors are commonly embedded in most of electronic personal devices like laptops, smart-phones or smart-watches. 
Typically, these sensors have been used to retrieve information like heart rate, blood oxygen level or user's fingerprint, for applications ranging from healthcare to biometric-based authentication. Previous studies \cite{ pulseid_article, pulse_zhang, pulse_spachos} have found that biometric signals obtained from sensors such as gyroscopes, accelerometers, or photoplethysmograms (PPG) can be processed to extract both further information and different patterns from those they were originally designed for. Furthermore and together with the advent of end-to-end neural network architectures, showing astonishing performances, 
they might translate into novel use cases and applications for biometrics in common wearable devices. Some recent works have studied the feasibility of using the PPG signal \cite{Challoner_1974} for user authentication, as shown in \cite{pulse_zhang}, \cite{pulse_spachos}. 
Typically, these methods involve both expert knowledge and considerable effort for the extraction of relevant biomarkers, which can encode user's characteristics. Recently, in \cite{pulseid_article}, it was shown that Convolutional Neural Network (CNN) methods can be used for automatically extracting such relevant features, proposing an end-to-end neural network architecture for the task of user's identity verification. The proposed fully end-to-end system, capable of automatically extract discriminant biomarkers, outperforms classical reported methods in both TROIKA \cite{troika_paper}  and self-recorded databases. 

Such findings motivate this work, which intends to extract relevant information from PPG signal and open up a door for further application possibilities, besides other recent proposed applications as user authentication. Some works have proven that speech information can be mined using indirect devices other than microphones, like a gyroscope \cite{gyroscope_article} or an accelerometer \cite{accelerometer_article}.
So, a PPG sensor could be a proper candidate for detecting speech events, as any other sensor present in smart devices. \

\begin{figure*}[!ht]
\begin{center}
\includegraphics[width=\textwidth,height=6.5cm]{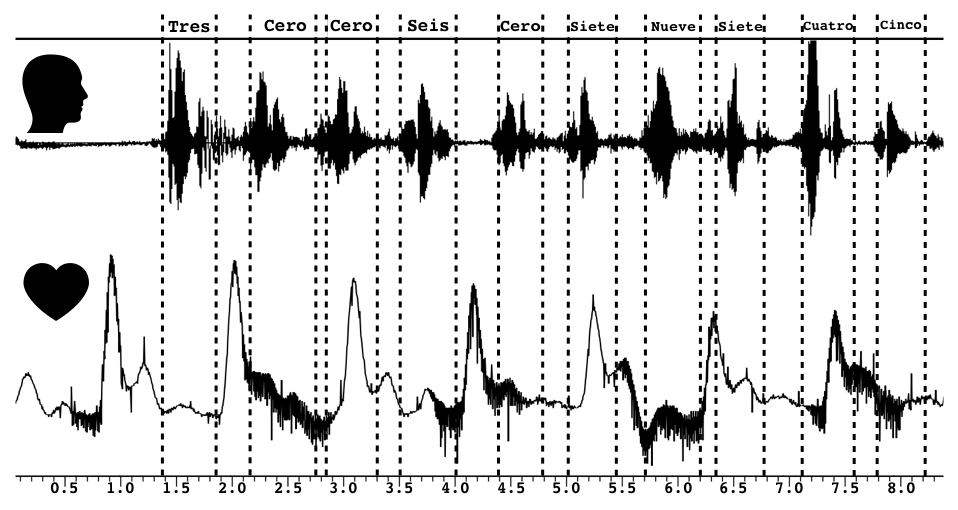}%
\end{center}
\caption{Multimodal 8-second recording of speaker $004$ from PulseID database. On top, spoken digits sequence, in the middle the audio waveform and, at the bottom the PPG signal. Dashed vertical lines depict the forced-alignment obtained between text and audio -- pronounced digits -- given by an automatic speech recognition system} 
\label{fig:pulse_example}
\end{figure*}


We hypothesize that speech produced by the user causes some sort of fluctuation in the PPG signal, either directly by low-frequency acoustic components from speech itself, or indirectly by the sensor movement \cite{gyroscope_article,accelerometer_article}. Some authors also have pointed that speech events might translate into PPG by an additional oxygen consumption \cite{moon2003two}. In any case, the detection of such 
fluctuations could suggest the existence of a speech event during a certain period of time. Wrapping up, the main aim of this work is to discern whether there is any correlation between speech/non-speech events and PPG.  Furthermore, our aim is to explore the feasibility to distinguish between different speech events as PPG serves to discriminate between users by detecting gender or identity. The latter would allow to develop tasks like speech/non-speech detection, speech enhancement or even word recognition, e.g., by augmenting current methods with PPG information.

\section{Previous work}
\label{sec:format}

Up to the authors' knowledge, no previous work has tried to find the correlation between PPG signal and speech, with or without deep learning methods. So, we will focus on the literature related to the topics converging in this work, such as speaker identification by using PPG signal, speech detection with unconventional devices, oxygen consumption on speech production or deep neural networks (DNN) applied to noise detection in electrocardiogram signal (ECG).

Some previous works \cite{pulse_zhang,pulse_spachos} studied time domain characteristics like time intervals, peaks, upward and downward slopes in PPG signal, used to infer a person's identity. In \cite{pulseid_article} it was shown that a CNN architecture allows for direct classification without feature extraction effort.
Further related works \cite{gyroscope_article} found out that gyroscopes capture acoustic vibrations from speech under frequencies of 200 Hz, which can be used to detect speech, identify speakers, or even parse such speech, using Short Time Fourier Transform (STFT), Mel-Frequency Cepstral Coefficients as features and by using machine learning classification methods like Support Vector Machine (SVM), Gaussian Mixture Model or Dynamic Time Warping. In a similar way, \cite{accelerometer_article} reported that acoustic vibrations leaked into an accelerometer signal as well, thus allowing for speech activity detection. They performed feature extraction using the Fast Fourier Transform (FFT), and various machine learning algorithms such as SVM or Na\"ive Bayes. Therefore, low frequencies from speech might leak as well into a PPG sensor, opening up for eavesdropping possibilities.

Similarly, noise detection task was done in \cite{john2018deep}, achieving the best results with a 16-layer CNN adapted from the VGG16 network \cite{simonyan2014very}. However, besides acoustic vibrations, an interesting indicator of speech might be the difference in oxygen consumption between speech and rest, measured by the PPG signal. In \cite{moon2003two} it was proven that oxygen consumption increases when the vocal effort (speech volume) is higher, and that a higher frequency of pronounced syllables yields a higher oxygen consumption as well. The first result reinforces the hypothesis that speech or non-speech events could be classified from a PPG signal because of the oxygen concentration in blood.

\section{Methodology}
\label{sec:pagestyle}




\subsection{Data source synchronization}
\label{ssec:datasync}
The PulseID dataset introduced in \cite{pulseid_article} is employed to assess the proposed CNN architectures. It was collected using the open source and low-cost photoplethysmograph sensor described in \cite{pulse_sensor}, which consists of a green LED and a photo-detector, sampling at a 200 Hz rate. For a sample of PPG signal, see Figure \ref{fig:pulse_example}. \

PPG waveform is synchronized up with audio recordings at same PPG recording device. For this purpose, a developed python code exploits threading capabilities of a Raspberry Pi 3 device, see \cite{pulseid_article,repo_guillem} for more details. The analysis of error measurements shows a tolerant sampling rate deviation ($\mu = 13.32\,\mu s$ and $\sigma = 202.58\,\mu s$) per subject, which lies within an acceptable margin for speech events. Audio signal is recorded at a 44.1 kHz sample rate and 16 bits per sample, but a downsampling to 8 kHz is done for further processing.


\subsection{Recording protocol}
\label{ssec:subhead}

Data acquisition implied the participation of 31 subjects (25 males and 6 females), with ages ranging from 22 to 55 years old. Five different types of speech recordings were performed through a microphone, including the pronunciation of certain words and sequence numbers like PINs. All the experiments include at least 30 seconds of pulse and audio signals, except for the fifth one, which took place during 1 minute. \

The first experiment involves subjects saying two random credit card numbers of 16 digits each at a regular pace, with a longer pause between both numbers. The second one contains a 30-second recording of the pulse and the audio without any speech from the subject. The third experiment, similar to the first one, includes subjects uttering four random 6-digit PIN numbers, also with a longer pause between two consecutive PIN numbers. The fourth experiment uses phonetically rich sentences commonly employed for ASR corpus, uttered by the subjects at a regular pace and with a longer pause between sentences. The last experiment includes one minute of free speech, where the subjects typically describe their environment \footnote{The {\it PulseID} dataset, including audio and alignments, is available upon request from the authors and agreement of EULA for research purposes.}. Both phonetically rich sentences and protocol guidelines are following those in \cite{van-den-heuvel-etal-2004-sala}. 






\subsection{Audio and text forced-alignment}
\label{ssec:align}
Once the PPG and the audio signals are synchronously recorded, the latter is used for getting speech timestamps at the frame level. The speech labelling is performed by forced-alignment of text sentences/numbers and audio. It is accomplished by an Automatic Speech Recognizer (ASR), previously trained with more than two hundred hours of Spanish speech and the Kaldi toolkit \cite{kaldi_article}. We make use of one of prototypes in \cite{prosodic_cues}.
Since the words pronounced are known in advance -- except for the free speech case, which is not annotated --- the ASR performs a lattice decoding and iterativelly refine it by seeking the phoneme sequence that maximizes a given word sequence. Nonetheless, previous methodology is prone to produce few errors, especially whenever strong background noise happens, so still a manual cleaning is done and non-accurate time stamps are corrected by hand.


\subsection{Neural network architectures}
\label{ssec:arc}
Five different Convolutional Neural Network (CNN) based architectures are benchmarked for speech detection, speaker identification and gender classification tasks: the CNN net from \cite{pulseid_article} -- from now on PulseNet -- plus two variants, the inverted VGG-like model employed for noise detection in ECG signals \cite{john2018deep} and a CNN which uses 2D STFT as an input -- CNN-2D --, adapted from a kernel in Kaggle's FAT 2019 competition \cite{fat2019}. 
The PulseNet initial kernel sizes for the three convolutional layers reported in \cite{pulseid_article}, which are $L_{1}$, $L_{2}$, $ L_{3}$ $= 50, 30, 20$, are optimized by brute force, exploring configurations of kernel sizes $L_{i} \in$ [2,3,4,...,180,200]. Using a PulseNet model with smaller kernel sizes aims at exploring shorter time correlations. From such explorations, we report the results on the best performing variants: PulseNet-Var1 ($L_{1}$, $L_{2}$, $ L_{3}$ $= 50, 10, 4$) and PulseNet-Var2 ($L_{1}$, $L_{2}$, $ L_{3}$ $= 15, 8, 2$). The code for training and testing these models is written in Python, mainly using PyTorch library \cite{paszke2017automatic}, and initially forked from the implementation in \cite{fonollosa}. The repository to replicate following experiments can be found at \cite{pulsespeech_repo}.


\begin{figure}[!t]%
\centering
\includegraphics[clip,width=\textwidth]{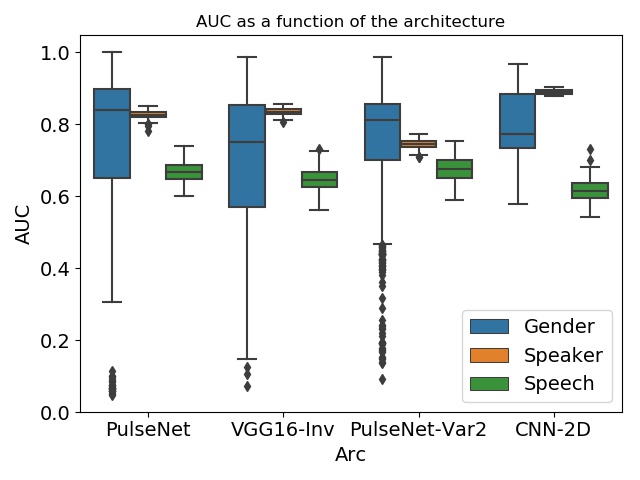} 
\caption{Boxplot of AUC values for test set evaluation given by four different architectures. Results are depicted across $100$ repetitions }%
\label{fig:architectures}%
\end{figure}




\begin{table*}[!ht]
\renewcommand{\arraystretch}{1.50}
\caption{Result scores for speech detection, gender detection and speaker verification test tasks, averaged over 100 repetitions, errors given by the standard error of the mean. PPG signals are split in smaller samples according to the window size column and fed to five different CNN architectures.}
\label{speech_nonspeech_final_table}
\small{
\begin{center}
\begin{tabular}{lccccccccc}
\hline
 & & \multicolumn{2}{c}{Speech Detection}  & & \multicolumn{2}{c}{Gender Detection} & & \multicolumn{2}{c}{Speaker Verification}   \\
\cline{3-4}\cline{6-7}\cline{9-10}
\bfseries Neural Network  &  \bfseries Window Size (s) & \bfseries AUC & \bfseries F1-Score & &  \bfseries AUC &  \bfseries F1-Score & & \bfseries AUC &  \bfseries F1-Score \\
\hline\hline
PulseNet [50,10,4] & 1 & $\boldsymbol{68.2 \pm 0.3}$ & 63.9 $\pm 0.3$ & & 77 $\pm 5$ & 61 $\pm 5$ & & 80.0 $\pm 0.1$ & 60.7 $\pm 0.2$  \\
PulseNet [15,8,2]  & 1 & 67.6 $\pm 0.3$ & 64.0 $\pm 0.3$ & & 75 $\pm 5$ & 60 $\pm 5$ & & 74.4 $\pm 0.1$ & 48.7 $\pm 0.3$ \\
PulseNet & 1 & 66.8 $\pm 0.3$ & 63.3 $\pm 0.3$ & & 74 $\pm 2$ & 60 $\pm 2$ & & 82.7 $\pm 0.1$ & 66.1 $\pm 0.3$ \\
VGG16-Inv & 1 & 64.6 $\pm 0.3$ & 62.6 $\pm 0.3$ & & 68 $\pm 2$ & 55 $\pm 2$ & & 83.4 $\pm 0.1$ & 67.3 $\pm 0.2$ \\
CNN-2D & 1 & 61.7 $\pm 0.3$ & 60.4 $\pm 0.3$ & & $\boldsymbol{79 \pm 1}$ & $\boldsymbol{66 \pm 1}$ & & $\boldsymbol{89.0 \pm 0.1}$ & $\boldsymbol{78.5 \pm 0.2}$ \\
PulseNet [50,10,4] & 0.4 & 66.7 $\pm 0.3$ & 64.0 $\pm 0.3$ & & 70 $\pm 2$ & 59 $\pm 2$ & & 68.4 $\pm 0.1$ & 37.9 $\pm 0.1$ \\
PulseNet & 0.4 & 67.6 $\pm 0.3$ & $\boldsymbol{64.7 \pm 0.3}$ & & 66 $\pm 2$ & 56 $\pm 2$ &  & 70.7 $\pm 0.1$ & 42.3 $\pm 0.1$ \\
\hline
\end{tabular}
\end{center}
}
\end{table*}


Cross-entropy loss function is used together with Stochastic Gradient Descent (SGD) optimizer. Batch size and learning rate are fine tuned for every architecture configuration to values allowing the model to safely decrease the loss whilst augmenting label accuracy. Particularly, a variant of SGD is implemented, which is known as SGD with Restarts (SGDR) \cite{loshchilov2016sgdr} and consists of decreasing the learning rate in the form of half a cosine curve, restarting it at the minimum to the original value.



\section{Experiments and results}
\label{sec:typestyle}

For every PPG recording, the signal is split with a rolling window, which is sliced along the PPG time axes. Every PPG window sample is marked as \emph{speech} or \emph{non-speech}, as indicated by audio timestamps (see section \ref{ssec:align}). 
Since there are short moments during speech where the speaker stops and breathes, if a \emph{speech} signal window contains more than a $2\%$ of silence, it is discarded, ensuring that the speaker was speaking the most of the time during signal samples labeled as \emph{speech}. After exploring several sizes and strides for the rolling window, the 1-second window as in \cite{pulseid_article} seems to yield the best results. Some experiments are done subdividing such 1-second windows in smaller windows of 0.4 seconds, each one of them overlapped in a $70\%$ with the adjacent ones, to explore the effect of overlapping during training. The decision on the final label of a 1 second sample in the test is done by summing the log-probabilities obtained from the forward pass of each 0.4-second sub-sample.

Every result score is averaged along 100 experiments, each one of them with shuffled training ($64\%$), validation ($16\%$) and testing ($20\%$) partitions. However, for the gender classification experiment the samples from 2 males and 2 females speakers are kept exclusively only for the testing set. 20 folds with different male-female combinations are tested, averaging the performance scores through 25 experiment repetitions with shuffled training and validation sets. Because of the bigger size in the CNN-2D network, the repetitions have been reduced to 10 times for a 10-fold in the gender detection task and 22 times for the speaker verification one, in order to save computational time. AUC and F1 weighted scores measure how good the model is, preferred over accuracy, in order to account for class imbalances, specially for speech detection ($33\%$/$66\%$ for speech/non-speech) and gender classification ($81\%$/$19\%$ for male/female).  

As can be seen in Table \ref{speech_nonspeech_final_table} and Figure \ref{fig:architectures}, the best results for speech detection are achieved by the PulseNet architectures, attaining a highest scoring mean AUC of $68.2 \pm 0.3\%$ with PulseNet [50,10,4]. This configuration seems to benefit of smaller kernel sizes than the original PulseNet to find correlations in the signal, suggesting that speech fluctuations could be really fine-grained. Using the original PulseNet with a smaller 0.4 s window sizes practically matches this result without need of reducing the kernels, which also points out in the direction of zooming in in the signal to find such correlations. Despite the CNN-2D architecture yields the worst result for speech detection, it achieves the best scores for gender detection and speaker verification tasks, with $79 \pm 1\%$ and $89.0 \pm 0.1\%$ AUCs respectively, surpassing PulseNet's performance particularly for the latter. The VGG16-Inv model shows fairly good performance at all tasks, but does not achieve any highest scoring result. It seems that mining speaker characteristics is better done with deeper CNNs with varying receptive fields at each filter, which is the case for CNN-2D and VGG16 architectures. Particularly, an approach with a STFT spectrogram like in CNN-2D might be better than using just the raw signal, as opposed to the speech detection task, which benefits from the smaller and parallel filters of PulseNet architectures. High variance in gender results is due to the low number of subjects, specially females. Depending on the speaker, the network achieves excellent or poor detection rates.

\section{Conclusion}
\label{sec:majhead}
In this work, we have explored end-to-end CNN architectures to detect speech and users' gender and identity characteristics. We have found that all five architectures used are able to distinguish speech from silence through PPG signals, not with a high accuracy, but indeed way over the random baseline, with all AUC and F1 scores over 60\%, averaged in runs of 100 experiments, and achieving almost a 70\% AUC with the best architecture. Such model is a variant of the original PulseNet configuration, with smaller kernel sizes, which suggests that speech information might appear in PPG signal as some sort of fine-grained fluctuation. Such results motivate further research, in order to clarify the nature of these fluctuations, and to improve the performance of existing systems, which would benefit of larger datasets and less noisy sensors. Furthermore, bi-dimensional CNN architectures using STFT spectrograms have shown to be the best option for extracting speaker characteristics like gender or identity, improving the performance of the PulseNet network for the latter task.

\vfill\pagebreak



\bibliographystyle{IEEEbib}
\bibliography{strings,refs}

\end{document}